\documentclass[aip,reprint]{revtex4-1}
\usepackage{float}
\usepackage{makeidx}
\usepackage{amsmath,amssymb,graphics,epsfig,color,times,bbm}

\begin{document}

\title{Experimental generation of tripartite polarization entangled states
of bright optical beams}
\author{Liang Wu}
\affiliation{State Key Laboratory of Quantum Optics and Quantum Optics Devices, Institute of Opto-Electronics, Shanxi University, Taiyuan, 030006, People's Republic of China}
\author{Zhihui Yan}
\affiliation{State Key Laboratory of Quantum Optics and Quantum Optics Devices, Institute of Opto-Electronics, Shanxi University, Taiyuan, 030006, People's Republic of China}
\affiliation{Collaborative Innovation Center of Extreme Optics, Shanxi University, Taiyuan 030006, People's Republic of China}
\author{Yanhong Liu}
\affiliation{State Key Laboratory of Quantum Optics and Quantum Optics Devices, Institute of Opto-Electronics, Shanxi University, Taiyuan, 030006, People's Republic of China}
\author{Ruijie Deng}
\affiliation{State Key Laboratory of Quantum Optics and Quantum Optics Devices, Institute of Opto-Electronics, Shanxi University, Taiyuan, 030006, People's Republic of China}
\author{Xiaojun Jia}
\email{jiaxj@sxu.edu.cn}
\affiliation{State Key Laboratory of Quantum Optics and Quantum Optics Devices, Institute of Opto-Electronics, Shanxi University, Taiyuan, 030006, People's Republic of China}
\affiliation{Collaborative Innovation Center of Extreme Optics, Shanxi University, Taiyuan 030006, People's Republic of China}
\author{Changde Xie}
\affiliation{State Key Laboratory of Quantum Optics and Quantum Optics Devices, Institute of Opto-Electronics, Shanxi University, Taiyuan, 030006, People's Republic of China}
\affiliation{Collaborative Innovation Center of Extreme Optics, Shanxi University, Taiyuan 030006, People's Republic of China}
\author{Kunchi Peng}
\affiliation{State Key Laboratory of Quantum Optics and Quantum Optics Devices, Institute of Opto-Electronics, Shanxi University, Taiyuan, 030006, People's Republic of China}
\affiliation{Collaborative Innovation Center of Extreme Optics, Shanxi University, Taiyuan 030006, People's Republic of China}

\begin{abstract}
The multipartite polarization entangled states of bright optical beams
directly associating with the spin states of atomic ensembles are one of the
essential resources in the future quantum information networks, which can be
conveniently utilized to transfer and convert quantum states across a
network composed of many atomic nodes. In this letter, we present the
experimental demonstration of tripartite polarization entanglement described
by Stokes operators of optical field. The tripartite entangled states of
light at the frequency resonant with D1 line of Rubidium atoms are
transformed into the continuous variable polarization entanglement among
three bright optical beams via an optical beam splitter network. The
obtained entanglement is confirmed by the extended criterion for
polarization entanglement of multipartite quantized optical modes.
\end{abstract}

\maketitle

Quantum entanglement plays the kernel role in the developing of quantum
information technology and has been applied in a variety of quantum
communication and computation protocols \cite%
{Brau,Jiaa12,Boumeester97,Boschi98,Furusawa08,Appel08,Polzik10,Su13}. At
present, the interest has focused on building the quantum internet \cite%
{Kimble08} composed of quantum nodes and quantum information transmission
channel \cite{Jia12,Chen14}. Continuous-variable (CV) polarization
entanglement of optical field can be manipulated and detected with high
efficiency and the bright polarization entangled beams can be directly
measured without the need of a local oscillator \cite%
{Polzik01,Lam02,Leuchs03,Timur12}. Furthermore, both the polarization of
light and atomic spin are described by Stokes operators, and the
fluctuations of the polarization variables can be easily mapped onto the
collective fluctuations of an atomic ensemble, thus the quantum state
transfer between CV polarization states and spin states of atomic ensembles
can be conveniently realized \cite{Polzik01}. In 2002, N. V. Korolkova et
al. introduced the physical concept about CV polarization entangled states
and proposed schemes of generating and characterizing them \cite{Korolkova02}%
. P. K. Lam's group experimentally demonstrated CV polarization squeezing
and bipartite entanglement by means of two degenerate optical parameter
amplifiers (DOPAs) \cite{Lam02}. Then the polarization entanglement between
two optical modes was realized by G. Leuchs's group with the asymmetric
fiber-optic Sagnac interferometer \cite{Leuchs03,Leuchs07}. The polarization
entanglement was also produced in cold cesium atoms placed inside an optical
cavity with high finesse \cite{Giacobino04}.

However, the bipartite entanglement is not enough to establish quantum
networks, thus we have to prepare the polarization entangled states with
more than two submodes. Here we report the experimental generation of CV
tripartite polarization entangled states of light resonant on the Rb D1 line
(795 nm), which are suitable for implementing optical storage and realizing
the interaction between light and atoms. According to the inseparability
criterion for the multipartite polarization entanglement deduced by us
before \cite{Yan},\ the obtained polarization entangled state is
characterized quantitatively. The experimentally produced tripartite
entangled states also satisfies the criterion for the genuine multipartite
entanglement deduced by R. Y. Teh and M. D. Reid \cite{Teh}. The
experimental system and scheme can be directly extended to produce CV
polarization entangled states with more submodes.

In quantum optics, the Stokes operators ($\hat{S}_{0}$, $\hat{S}_{1}$, $\hat{%
S}_{2}$ and $\hat{S}_{3}$) are usually used to describe the polarization
state of light \cite{Stokes}, which satisfy a spherical equation $\hat{S}%
_{1}^{2}+$ $\hat{S}_{2}^{2}+$ $\hat{S}_{3}^{2}=$ $\hat{S}_{0}^{2}+2\hat{S}%
_{0}$ and constitute a Poincar\'{e} sphere\cite{Timur12,Korolkova02}. Where $%
\hat{S}_{0}$ represents the beam intensity whereas $\hat{S}_{1}$, $\hat{S}%
_{2}$ and $\hat{S}_{3}$ characterize its polarization and form a Cartesian
axes system, which can be easily mapped to the spin operators of the atomic
media. The Stokes parameters for pure states can be described by the
corresponding annihilation $\hat{a}_{H(V)}$ and creation $\hat{a}%
_{H(V)}^{\dagger }$ operators of the constituent horizontally (subscript H)
and vertically (subscript V) polarized modes in the frequency space, that
are:

\begin{eqnarray}
\hat{S}_{0} &=&\hat{a}_{H}^{\dagger }\hat{a}_{H}+\hat{a}_{V}^{\dagger }\hat{a%
}_{V},\hat{S}_{2}=\hat{a}_{H}^{\dagger }\hat{a}_{V}e^{i\theta }+\hat{a}%
_{V}^{\dagger }\hat{a}_{H}e^{-i\theta }, \\
\hat{S}_{1} &=&\hat{a}_{H}^{\dagger }\hat{a}_{H}-\hat{a}_{V}^{\dagger }\hat{a%
}_{V},\hat{S}_{3}=i\hat{a}_{V}^{\dagger }\hat{a}_{H}e^{-i\theta }-i\hat{a}%
_{H}^{\dagger }\hat{a}_{V}e^{i\theta }.  \label{1}
\end{eqnarray}%
Where $\theta $ is the relative phase between the H and V-polarization modes.

\begin{figure}[tbp]
\centerline{
\includegraphics[width=75mm]{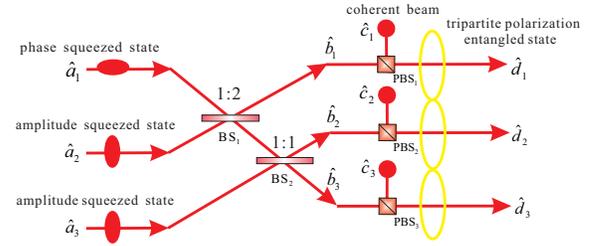}
}
\caption{Schematic for the generation of tripartite polarization entangled
state.}
\end{figure}

The schematic for the generation of tripartite polarization entangled state
is shown in Fig. 1. A quadrature phase squeezed state of light ($\hat{a}_{1}$%
) and two quadrature amplitude squeezed state of light ($\hat{a}_{2}$ and $%
\hat{a}_{3}$) interfere on a beam splitters BS$_{1}$ (BS$_{2}$), with the
ratio of\ reflectivity and transmissivity R:T=1:2 (1:1) to generate a
tripartite GHZ-like entangled state ($\hat{b}_{1}$, $\hat{b}_{2}$ and $\hat{b%
}_{3}$) \cite{Furusawa03}. The three submodes $\hat{b}_{1}$, $\hat{b}_{2}$
and $\hat{b}_{3}$, each of which is a weak horizonal polarized (H) states,
are coupled with three strong vertical polarized (V) coherent beams $\hat{c}%
_{1}$, $\hat{c}_{2}$ and $\hat{c}_{3}$ on three polarization beam splitters
(PBS$_{1-3}$), respectively. For simplicity and without loss of generality,
the average power of the three squeezed states of light (coherent light) is
adjusted to equal, that is: $\alpha _{a_{1}}^{2}=$ $\alpha
_{a_{2}}^{2}=\alpha _{a_{3}}^{2}=\alpha _{a}^{2}$ ($\alpha _{c_{1}}^{2}$ $%
=\alpha _{c_{2}}^{2}=\alpha _{c_{3}}^{2}=$ $\alpha _{c}^{2}$). The ratio of
the intensity of the squeezing and coherent light is: $\alpha
_{a}^{2}/\alpha _{c}^{2}=1/30$. The phase difference $\theta $ on the beam
splitters BS$_{1,2}$ and PBS$_{1-3}$ all are controlled to $2k\pi $ ($k$ is
an integer). The noise operators of the quadrature amplitude $\delta \hat{X}%
_{a_{i}}^{+}(\Omega )$ (phase $\delta \hat{X}_{a_{i}}^{-}(\Omega )$) of a
squeezed state at the sideband frequency ($\Omega $) can be expressed as $%
\delta \hat{X}_{a_{2(3)}}^{+}(\Omega )=e^{-r_{2(3)}}\delta \hat{X}%
_{a_{2(3)}}^{+(0)}(\Omega )$, $\delta \hat{X}_{a_{2(3)}}^{-}(\Omega
)=e^{r_{2(3)}+r_{2(3)}^{\prime }}\delta \hat{X}_{a_{2(3)}}^{-(0)}(\Omega )$ (%
$\delta \hat{X}_{a_{1}}^{+}(\Omega )=e^{r_{1}+r_{1}^{\prime }}\delta \hat{X}%
_{a_{1}}^{+(0)}(\Omega )$, $\delta \hat{X}_{a_{1}}^{-}(\Omega
)=e^{-r_{1}}\delta \hat{X}_{a_{1}}^{-(0)}(\Omega )$), where $\delta \hat{X}%
_{a_{i}}^{\pm (0)}(\Omega )$ are the amplitude ($+$) and phase ($-$) noise
operators of the input beams of DOPAs, $r_{i}$ is the squeezing parameter
and $r_{i}^{\prime }$ is the factor of extra noise on the antisqueezed
quadrature components \cite{Zhang03}. Since $\alpha _{a}^{2}\ll \alpha
_{c}^{2}$ the quantum fluctuation variance of Stokes operators are expressed
as

\begin{eqnarray}
&&\delta ^{2}\hat{S}_{0_{d_{1}(d_{2},d_{3})}}(\Omega )=\delta ^{2}\hat{S}%
_{1_{d_{1}(d_{2},d_{3})}}(\Omega )=4\alpha _{c}^{2}\delta ^{2}\hat{X}%
_{c_{1(2,3)}}^{+}(\Omega ),  \notag \\
&&\delta ^{2}\hat{S}_{2_{d_{1}}}(\Omega )=4\alpha _{c}^{2}(\frac{%
e^{2r_{1}+2r_{1}^{^{\prime }}}}{3}\delta ^{2}\hat{X}_{a_{1}}^{+(0)}(\Omega )+%
\frac{2e^{-2r_{2}}}{3}\delta ^{2}\hat{X}_{a_{2}}^{+(0)}(\Omega )),  \notag \\
&&\delta ^{2}\hat{S}_{3_{d_{1}}}(\Omega )=4\alpha _{c}^{2}(\frac{e^{-2r_{1}}%
}{3}\delta ^{2}\hat{X}_{a_{1}}^{-(0)}(\Omega )+\frac{2e^{2r_{2}+2r_{2}^{%
\prime }}}{3}\delta ^{2}\hat{X}_{a_{2}}^{-(0)}(\Omega )),  \notag \\
&&\delta ^{2}\hat{S}_{2_{d_{2}}}(\Omega )=4\alpha _{c}^{2}(\frac{%
e^{2r_{1}+2r_{1}^{\prime }}}{3}\delta ^{2}\hat{X}_{a_{1}}^{+(0)}(\Omega )
\notag \\
&&-\frac{e^{-2r_{2}}}{6}\delta ^{2}\hat{X}_{a_{2}}^{+(0)}(\Omega )+\frac{%
e^{-2r_{3}}}{2}\delta ^{2}\hat{X}_{a_{3}}^{+(0)}(\Omega )),  \notag \\
&&\delta ^{2}\hat{S}_{3_{d_{2}}}(\Omega )=4\alpha _{c}^{2}(\frac{e^{-2r_{1}}%
}{3}\delta ^{2}\hat{X}_{a_{1}}^{-(0)}(\Omega )  \notag \\
&&-\frac{e^{2r_{2}+2r_{2}^{\prime }}}{6}\delta ^{2}\hat{X}%
_{a_{2}}^{-(0)}(\Omega )+\frac{e^{2r_{3}+2r_{3}^{\prime }}}{2}\delta ^{2}%
\hat{X}_{a_{3}}^{-(0)}(\Omega )),  \notag \\
&&\delta ^{2}\hat{S}_{3_{d_{3}}}(\Omega )=4\alpha _{c}^{2}(\frac{%
e^{2r_{1}+2r_{1}^{\prime }}}{3}\delta ^{2}\hat{X}_{a_{1}}^{+(0)}(\Omega )
\notag \\
&&-\frac{e^{-2r_{2}}}{6}\delta ^{2}\hat{X}_{a_{2}}^{+(0)}(\Omega )-\frac{%
e^{-2r_{3}}}{2}\delta ^{2}\hat{X}_{a_{3}}^{+(0)}(\Omega )),  \notag \\
&&\delta ^{2}\hat{S}_{2_{d_{3}}}(\Omega )=4\alpha _{c}^{2}(\frac{e^{-2r_{1}}%
}{3}\delta ^{2}\hat{X}_{a_{1}}^{-(0)}(\Omega )  \notag \\
&&-\frac{e^{2r_{2}+2r_{2}^{\prime }}}{6}\delta ^{2}\hat{X}%
_{a_{2}}^{-(0)}(\Omega )-\frac{e^{2r_{3}+2r_{3}^{\prime }}}{2}\delta ^{2}%
\hat{X}_{a_{3}}^{-(0)}(\Omega )).
\end{eqnarray}%
Where $\delta ^{2}\hat{S}_{j_{d_{k}}}(\Omega )$ ($j=0,1,2,3$. $k=1,2,3)$ are
the variances of Stokes operators of beam $d_{k}$, $\delta ^{2}\hat{X}%
_{c_{1(2,3)}}^{+}(\Omega )$ are the variances of quadrature amplitude
operators of beam $c_{1-3}$.

W. P. Bowen et al. extended the inseparability criterion characterizing CV
quadrature entanglement \cite{Duan,Simon} to CV bipartite polarization
entanglement \cite{Lam02}. P. van Loock and A. Furusawa gave the
inseparability criterion formula for multipartite states in 2003 \cite%
{Furusawa03}. Very recently, we deduced the tripartite inseparability
criterion of Stokes operators for optical beams \cite{Yan} based on the
theoretical analysis in Ref. [22]:

\begin{eqnarray}
I_{1} &\equiv &\frac{\delta ^{2}(\hat{S}_{2_{d_{2}}}-\hat{S}%
_{2_{d_{3}}})+\delta ^{2}(g_{1}\hat{S}_{3_{d_{1}}}+\hat{S}_{3_{d_{2}}}+\hat{S%
}_{3_{d_{3}}})}{4\left\vert \alpha _{c}^{2}-\alpha _{a}^{2}\right\vert }\geq
1,  \notag \\
I_{2} &\equiv &\frac{\delta ^{2}(\hat{S}_{2_{d_{1}}}-\hat{S}%
_{2_{d_{3}}})+\delta ^{2}(\hat{S}_{3_{d_{1}}}+g_{2}\hat{S}_{3_{d_{2}}}+\hat{S%
}_{3_{d_{3}}})}{4\left\vert \alpha _{c}^{2}-\alpha _{a}^{2}\right\vert }\geq
1,  \notag \\
I_{3} &\equiv &\frac{\delta ^{2}(\hat{S}_{2_{d_{1}}}-\hat{S}%
_{2_{d_{2}}})+\delta ^{2}(\hat{S}_{3_{d_{1}}}+\hat{S}_{3_{d_{2}}}+g_{3}\hat{S%
}_{3_{d_{3}}})}{4\left\vert \alpha _{c}^{2}-\alpha _{a}^{2}\right\vert }\geq
1.  \notag \\
&&
\end{eqnarray}%
Where, $I_{1}$, $I_{2}$ and $I_{3}$ are the normalized correlation variances
among Stokes operators, $g_{j}$ ($j=1,2,3$) are the adjustable classical
gains for minimizing the correlation variances. When any two in the three
inequalities are simultaneously violated, the three optical modes are in a
tripartite polarization inseparable state.

In order to show the dependence of the correlation variances on the
experimental parameters, the expressions of the normalized tripartite
correlation variances based on the experimental parameters of three optical
submodes can be obtained:

\begin{eqnarray}
I_{1} &=&\{\alpha
_{c}^{2}[12e^{-2r_{3}}+2(g_{1}+2)^{2}e^{-2r_{1}}+4(g_{1}-1)^{2}  \notag \\
&&e^{2(r_{2}+r_{2}^{\prime })}]\}/(24\left\vert \alpha _{c}^{2}-\alpha
_{a}^{2}\right\vert ),  \notag \\
I_{2} &=&\{\alpha
_{c}^{2}[3e^{-2r_{3}}+9e^{-2r_{2}}+2(g_{2}+2)^{2}e^{-2r_{1}}+3(g_{2}-1)^{2}
\notag \\
&&e^{2(r_{3}+r_{3}^{\prime })}+(g_{2}-1)^{2}e^{2(r_{2}+r_{2}^{\prime
})}]\}/(24\left\vert \alpha _{c}^{2}-\alpha _{a}^{2}\right\vert ),  \notag \\
I_{3} &=&\{\alpha
_{c}^{2}[3e^{-2r_{3}}+9e^{-2r_{2}}+2(g_{3}+2)^{2}e^{-2r_{1}}+3(g_{3}-1)^{2}
\notag \\
&&e^{2(r_{3}+r_{3}^{\prime })}+(g_{3}-1)^{2}e^{2(r_{2}+r_{2}^{\prime
})}]\}/(24\left\vert \alpha _{c}^{2}-\alpha _{a}^{2}\right\vert ).
\end{eqnarray}

Calculating the minimum values of the expressions (5), we get the dependence
of the optimal gains ($g_{i}^{opt}$) on the experimental parameters, that
are:
\begin{eqnarray}
g_{1}^{opt} &=&\frac{2e^{2r_{1}+2r_{2}+2r_{2}^{\prime }}-2}{%
2e^{2r_{1}+2r_{2}+2r_{2}^{\prime }}+1}, \\
g_{2}^{opt} &=&g_{3}^{opt}=\frac{e^{2r_{1}+2r_{2}+2r_{2}^{\prime
}}+3e^{2r_{1}+2r_{3}+2r_{3}^{\prime }}-4}{e^{2r_{1}+2r_{2}+2r_{2}^{\prime
}}+3e^{2r_{1}+2r_{3}+2r_{3}^{\prime }}+2}.
\end{eqnarray}

In 2014, R. Y. Teh and M. D. Reid pointed out the difference between the
genuine N-partite entanglement and full N-partite inseparability and
presented the criterion inequalities for the genuine multipartite
entanglement among optical modes\cite{Teh}. Based on Ref. [20] and [26], we
know that the sum of variances of an observable cannot be less than the
weighted sum of the variances of the component states for any mixture:

\begin{equation}
\delta ^{2}(\hat{S}_{2})+\delta ^{2}(\hat{S}_{3})\geq \Sigma
_{k}P_{k}(\delta _{k}^{2}(\hat{S}_{2})+\delta _{k}^{2}(\hat{S}_{3})),
\end{equation}%
where $P_{k}$ is a probability the system is separable across the
bipartition $k$ (thus, $\Sigma _{k}P_{k}=1$) and $\delta _{k}^{2}(\hat{S}%
_{2(3)})$ denotes the variance of $\hat{S}_{2(3)}$ for the system in the
state $\rho _{k}$\cite{Hof}. For tripartite state, since $I_{1}$ is the sum
of two variances, we can get

\begin{eqnarray}
I_{1} &\geq &P_{1}I_{1,1}+P_{2}I_{1,2}+P_{3}I_{1,3}  \notag \\
&\geq &P_{1}I_{1,1}+P_{2}I_{1,2}\geq P_{1}+P_{2}.
\end{eqnarray}

Similarly, $I_{2}\geq P_{2}+P_{3}$ and $I_{3}\geq P_{3}+P_{1}$. Since $%
\Sigma _{k}P_{k}=1$, for any mixture it must be true:

\begin{equation}
I_{1}+I_{2}+I_{3}\geq 2.
\end{equation}

That is, for the genuine tripartite polarization entanglement, the above
inequality must be violated\cite{Teh}.

\begin{figure}[tbp]
\centerline{
\includegraphics[width=90mm]{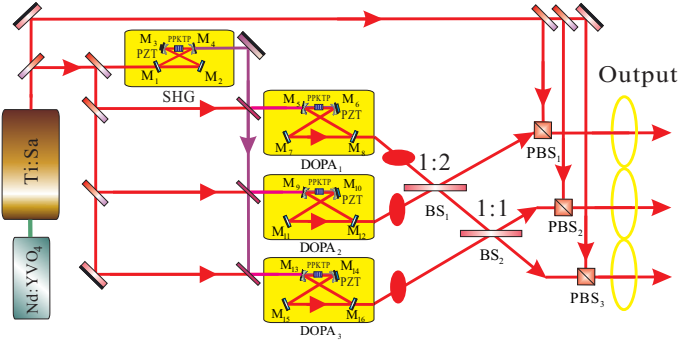}
}
\caption{The experimental setup for the generation of tripartite
polarization entangled state. Nd:YVO$_{4}$: Nd: YVO$_{4}$ green laser;
Ti:Sa, Titanium sapphire laser; SHG: second harmonic generation cavity;
DOPA: degenerate optical parameter amplifiers; BS$_{1-2}$: beam splitter
with different reflectivity; PBS$_{1-3}$: polarization beam splitter.}
\end{figure}
The experimental setup for the generation of tripartite polarization
entanglement is shown in Fig. 2. The Nd: YVO$_{4}$ green laser (DPSS
FG-VIIIB produced by Yuguang company) is used to pump the Titanium sapphire
laser (MBR 110 produced by the Coherent company). The output of the Titanium
sapphire laser is divided into three parts, the first part is used as the
seed beams of three DOPAs, the second part is used as the coherent beam for
transferring the quadrature entanglement to the polarization entanglement,
the rest is sent to the second harmonic generation (SHG) cavity for
obtaining the pump light of three DOPAs. The cavity for SHG is a four-mirror
ring cavity consisting of two plane mirrors (M$_{1}$, M$_{2}$), two
spherical mirrors (M$_{3}$, M$_{4}$) and a type-I phase matching $1\times
2\times 10mm^{3}$ PPKTP crystal \cite{Wen14,Deng13}. The plane mirror M$_{1}$
is used as the input coupler, which is coated with transmissivity of $13\%$
at 795 nm. The other three mirrors (M$_{2-4}$) are highly reflecting for
subharmonic optical field (795 nm) and M$_{4}$ is also coated with
anti-reflecting at 397.5 nm to be the output coupler\ of the second harmonic
optical field. Piezoelectric transducer (PZT) mounted on M3 and
Pound-Drever-Hall technique is used to lock the cavity length. The output
beams from SHG with wavelength at 397.5 nm are split to three parts to pump
three DOPAs.

The three DOPAs have the same configuration, which are also the four-mirror
ring cavity consisting of two plane mirrors and two spherical mirrors with
the radius curvature of 100 mm. A type-I phase matching PPKTP crystal, which
is in the middle of two spherical mirrors, is placed in a copper-made oven
and temperature-controlled around $54^{\circ }C$ with a peltier element for
achieving the optimal phase matching. The spherical mirrors (M$_{5}$, M$_{9}$
and M$_{13}$) are used as the input coupler of the corresponding DOPAs,
which are coated with anti-reflection for the pump field and
high-reflectivity for the subharmonic optical field. Another spherical
mirrors (M$_{6}$, M$_{10}$ and M$_{14}$) and the plane mirrors (M$_{7}$, M$%
_{11}$ and M$_{15}$) are coated with high reflection for the subharmonic
modes. Another plane mirrors (M$_{8}$, M$_{12}$ and M$_{16}$) coated with T
= 5.0\% for 795 nm are used as the output couplers. M$_{6}$ (M$_{10}$, M$%
_{14}$) is mounted on a PZT to scan actively the cavity length of the DOPA$%
_{1}$ (DOPA$_{2}$, DOPA$_{3}$) or lock it on resonance with injected seed
beam as needed by the Pound-Drever-Hall technique \cite{Jiaa12,Zhou15} . In
order to reduce the threshold and maximize the nonlinear conversion
efficiency in DOPAs, a beam waist of 39 $\mu m$ for the subharmonic optical
field is chosen by control the length of the cavity. The finesses of DOPA$%
_{1}$, DOPA$_{2}$ and DOPA$_{3}$ for the subharmonic mode are 111, 110 and
110, respectively.

\begin{figure}[tbp]
\centerline{
\includegraphics[width=75mm]{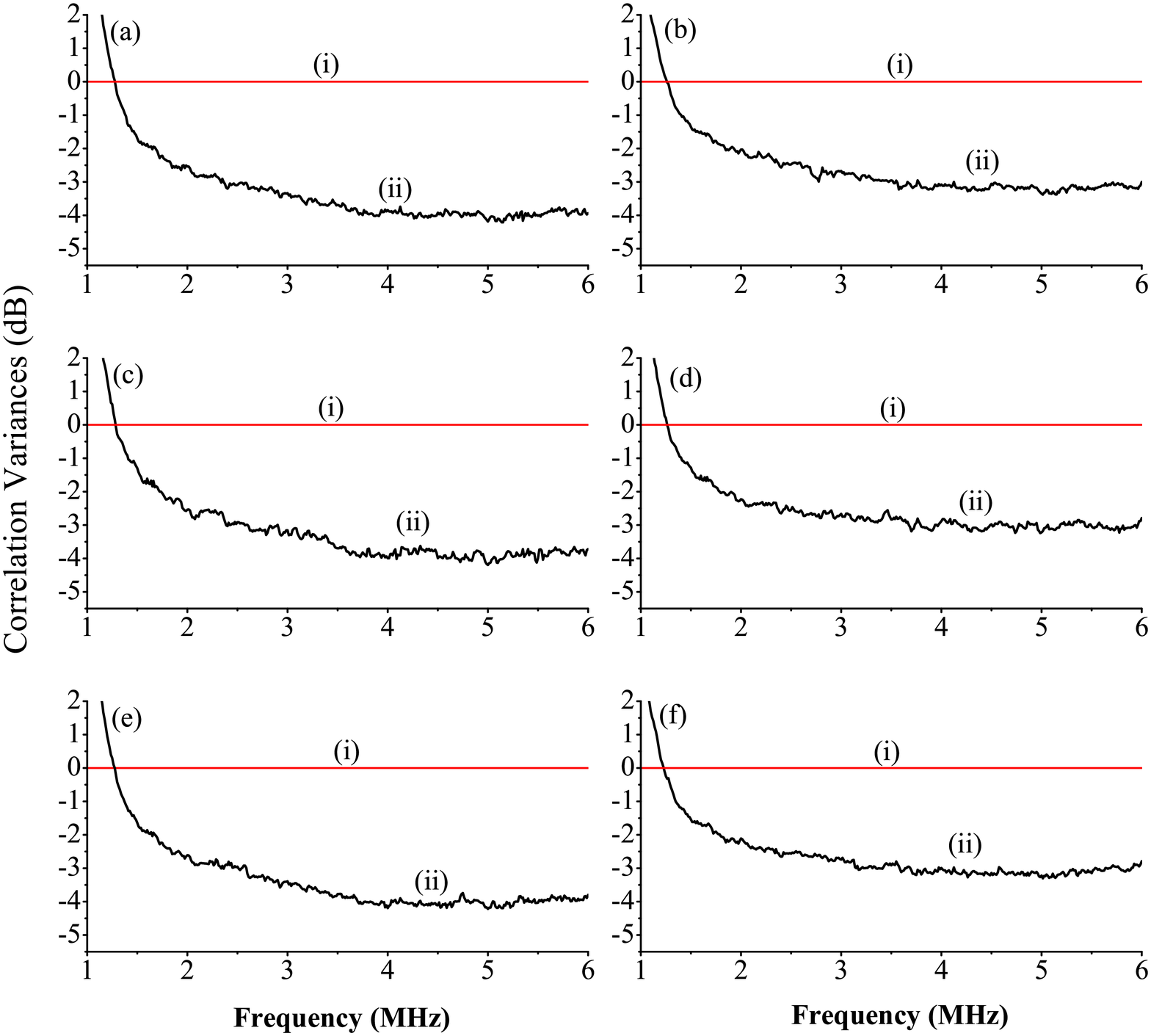}
}
\caption{The measured correlation variances of $\protect\delta ^{2}(\hat{S}%
_{2_{d_{2}}}-\hat{S}_{2_{d_{3}}})$ (a), $\protect\delta ^{2}(g_{1}\hat{S}%
_{3_{d_{1}}}+\hat{S}_{3_{d_{2}}}+\hat{S}_{3_{d_{3}}})$ (b), $\protect\delta %
^{2}(\hat{S}_{2_{d_{1}}}-\hat{S}_{2_{d_{3}}})$ (c), $\protect\delta ^{2}(%
\hat{S}_{3_{d_{1}}}+g_{2}\hat{S}_{3_{d_{2}}}+\hat{S}_{3_{d_{3}}})$ (d), $%
\protect\delta ^{2}(\hat{S}_{2_{d_{1}}}-\hat{S}_{2_{d_{2}}})$ (e) and $%
\protect\delta ^{2}(\hat{S}_{3_{d_{1}}}+\hat{S}_{3_{d_{2}}}+g_{3}\hat{S}%
_{3_{d_{3}}})$ (f) over the analysis frequency range from 1.0 to 6.0 MHz.
The trace (i) is the corresponding normalized SNL and the trace (ii) is the
quantum correlation noise. The measurement parameters of SA: Resolution
Bandwidth (RBW): 300kHz; Video Bandwidth (VBW): 300Hz.}
\end{figure}

When the relative phase between the pump field and seed field is locked to $%
(2k+1)\pi $ ($2k\pi $) ( $k$ is an integer), the output optical field from
DOPAs is quadrature amplitude (phase) squeezed state of light \cite{Mehmet11}%
. In our experiment DOPA$_{1}$ is locked to $2k\pi $, as well as DOPA$_{2}$
and DOPA$_{3}$ are locked to $(2k+1)\pi $ to obtain the needed quadrature
squeezed states as shown in Fig. 1. the power of pump beam for all three
DOPAs is about $40$ mW, and the power of the seed beam for the DOPA$_{1}$
and DOPA$_{2(3)}$ is about $0.2$ mW and $2$ mW respectively. In this case,
the power of the output beam from the three DOPAs is almost the same. The
quadrature phase squeezed state of light generated by DOPA$_{1}$ and the
quadrature amplitude squeezed state of light generated by DOPA$_{2}$ are
interfered on beam splitter BS$_{1}$. One of two output beams from BS$_{1}$
and the output of DOPA$_{3}$ are interfered on BS$_{2}$. The relative phase
between the two input beams of BS$_{1}$ and BS$_{2}$ is $2k\pi $. The
outcoming three optical beams from the two beam splitters are in a
tripartite GHZ-like entangled states. Then, the obtained tripartite
quadrature entangled states are transformed into tripartite polarization
entanglement by coupling with three strong coherent beams on PBS$_{1-3}$
with the phase difference of $2k\pi $. The three output beams are detected
by three sets of Stokes parameters measurement systems with a Spectrum
analyzer (SA), which have been introduced in the previous experiment \cite%
{Lam02}. The output from DOPA is a broadband quadrature squeezed state of
light, i.e. we can observe the squeezing phenomenon within the frequency
bandwidth of the DOPA. Since in the region of lower frequencies the quantum
noise of the laser is far higher than the shot noise limit (SNL) due to the
influence of the extra noises in the pump laser, we measure the correlation
variances over the analysis frequency range from 1.0 to 6.0 MHz. Figs. 3
show the measured correlation variances of $\delta ^{2}(\hat{S}_{2_{d_{2}}}-%
\hat{S}_{2_{d_{3}}})$, $\delta ^{2}(g_{1}\hat{S}_{3_{d_{1}}}+\hat{S}%
_{3_{d_{2}}}+\hat{S}_{3_{d_{3}}})$, $\delta ^{2}(\hat{S}_{2_{d_{1}}}-\hat{S}%
_{2_{d_{3}}})$, $\delta ^{2}(\hat{S}_{3_{d_{1}}}+g_{2}\hat{S}_{3_{d_{2}}}+%
\hat{S}_{3_{d_{3}}})$, $\delta ^{2}(\hat{S}_{2_{d_{1}}}-\hat{S}_{2_{d_{2}}})$
and $\delta ^{2}(\hat{S}_{3_{d_{1}}}+\hat{S}_{3_{d_{2}}}+g_{3}\hat{S}%
_{3_{d_{3}}})$, respectively. The traces (ii) are the measured quantum
correlation noises and the traces (i) are the corresponding normalized SNL.
When the correlation variances of $\delta ^{2}(g_{1}\hat{S}_{3_{d_{1}}}+\hat{%
S}_{3_{d_{2}}}+\hat{S}_{3_{d_{3}}}) $, $\delta ^{2}(\hat{S}_{3_{d_{1}}}+g_{2}%
\hat{S}_{3_{d_{2}}}+\hat{S}_{3_{d_{3}}})$ and $\delta ^{2}(\hat{S}%
_{3_{d_{1}}}+\hat{S}_{3_{d_{2}}}+g_{3}\hat{S}_{3_{d_{3}}})$ are measured ,
the optimal gains are chosen to minimize the correlation variances for
maximizing the measured entanglement. The experimentally optimal gains are
in good agreement with the theoretically calculated values ($%
g_{1}^{opt}=g_{2}^{opt}=g_{3}^{opt}=0.845$) from Eqs. (6-7). Both of the
correlation variances are below the corresponding SNL throughout the
frequency range from 1.3 MHz to 6.0 MHz. The best entanglement is observed
at 5 MHz with $I_{1}=0.42\pm 0.08$, $I_{2}=0.41\pm 0.08$, $I_{3}=0.42\pm
0.08 $ and $I_{1}+I_{2}+I_{3}=1.25\pm 0.07$, which violate the criteria for
both tripartite inseparability and genuine tripartite entanglement, thus we
say that the genuine tripartite polarization entanglement is verified.

To the conclusion, we present the experimental generation of CV multipartite
polarization entanglement by means of transforming the quadrature
entanglement into a polarization basis. In the presented scheme,
multipartite quadrature entangled states are the basic sources for the
generation of multipartite polarization entangled states. Although we just
combined a tripartite entangled state and three bright coherent optical
beams to produce a tripartite polarization entangled state in this
experiment, the entangled states with much more submodes are possible to be
produced if the multipartite quadrature entangled states are available.
Using the quadrature entangled states involving more submodes \cite%
{Su13,Yuka08}and a proper beamsplitter network, the presented method can be
extended to prepare polarization entangled states with more submodes and
thus has potential applications in the future quantum information networks.

We acknowledge the support from the Natural Science Foundation of China
(Grants Nos. 11322440, 11474190, 11304190,11504220), FOK YING TUNG Education
Foundation, Natural Science Foundation of Shanxi Province (Grant No.
2014021001).

\end{document}